# *Environment and Planning B* as a Journal:

# The interdisciplinarity of its environment and the citation impact

Loet Leydesdorff

Amsterdam School of Communications Research (ASCoR), University of Amsterdam

Kloveniersburgwal 48, 1012 CX  Amsterdam, The Netherlands.

loet@leydesdorff.net; http://www.leydesdorff.net

**Abstract**

The citation impact of *Environment and Planning B* can be visualized using its citation relations with journals in its environment as the links of a network. The size of the nodes is varied in correspondence to the relative citation impact in this environment. Additionally, one can correct for the effect of within-journal "self"-citations. The network can be partitioned and clustered using algorithms from social network analysis. After transposing the matrix in terms of rows and columns, the citing patterns can be mapped analogously. Citing patterns reflect the activity of the community of authors who publish in the journal, while being cited indicates reception. *Environment and Planning B* is cited across the interface between the social sciences and the natural sciences, but its authors cite almost exclusively from the domain of the *Social Science Citation Index*.

**Introduction**

The metaphor of the sciences as a landscape organized in different disciplines and specialties is an old one, but the idea that this landscape could perhaps be mapped by using aggregated journal-journal relations originated in the mind of the historian of science Derek de Solla Price when he was allowed access to the first experimental tape of the *Science Citation Index* 1961 (Price, 1965; Garfield & Sher, 1963). Price (1961) had been fascinated with the growth of scientific literature and its organization into journal structures ever since his study of the *Philosophical Transactions of the Royal Society* from its very beginning in 1665 (Price, 1951).[1]

Crane (1969, 1972) argued that the sciences are organized in relatively small communities which are controlled by invisible elites; Zuckerman & Merton (1971) emphasized the importance of the peer-review system in organizing these communities. Rewards in the community can be controlled in terms of reputations (Whitley, 1984). These self-reinforcing structures can be expected to lead to highly codified communications with strong boundaries (Leydesdorff & Van den Besselaar, 1997).

Might it be possible to operationalize sociological questions about interreading communities in terms of citation structures? (Mullins *et al*., 1977; Van den Besselaar & Leydesdorff, 1996). Soon after the publication of the first report on science indicators by the National Science Board of the U.S.A. in 1974, a conference was held where some of these founding fathers of the sociology of science met and discussed the potential of the *Science Citation Index* as a new source of scientific scholarship in terms of 'a metric of science' (Elkana, 1978). In 1978, the journal *Scientometrics* was launched for this very purpose and in the preface to the first issue, Price (1978) noted the preferential status of scientific literature as a source for such an enterprise:

> "[…] I feel that scientometrics has potentially an even greater possibility of success than econometrics or sociometrics or even general bibliometrics. It becomes apparent, even from our first few decades of analysis, that science and scientific activity is peculiarly measurable *and* peculiarly regular in its behavior even compared with other modes of scholarship." (Price, 1978, at p. 8)

---

[1] The first scientific journals were the *Philosophical Transactions of the Royal Society* published in 1665 and the French *Journal des Sçavants* shortly thereafter (cf. Leydesdorff, 1998).

Has Price's dream come true? Has scientometrics become a robust science? Is a universal mapping of science feasible? (Wouters & Leydesdorff, 1994). During the 1980s various attempts were made to generate a so-called *World Atlas of Science* from citation data. The Institute of Scientific Information (ISI) first published an atlas of science in 1981 (Garfield, 1981). This atlas was built on the pioneering work of Small & Griffith (1974) who used co-citations as links for mapping the sciences (Small & Greenley, 1985; Small *et al*., 1985; Small, 1999).

Garfield's (1972) original idea to map the sciences using journals as units of analysis had been taken up by a competing research team at Computer Horizons Inc. under the directorship of Francis Narin (Narin, 1972; Carpenter & Narin, 1973; Pinski & Narin, 1976). However, both these attempts failed because the relational or graph-analytic and the positional or factor-analytic approach were not sufficiently distinguished at the time (Burt, 1982; Leydesdorff, 1987 and 2006). The aim was to indicate both structures and hierarchies using a single representation summarizing the available information.

Evaluations of papers, authors, or journals presume a hierarchical model in which one can measure the standing of these units. Hierarchies are constructed relationally and can be mapped using trees or—in the graphical representation—dendograms. However, units of science, like specialties and disciplines develop concurrently and competitively; the mode of organization is mainly *heterarchical* and based on functional differentiation instead of stratification. In other words, networks are constructed relationally, but the architecture develops a structure in which the units are also *positioned*. Factor analysis enables us to reconstruct these positions in terms of the latent eigenvectors that span the network, while graph-analytical approaches focus on the vectors of observable relations. Because the subsystems are nested and the system is evolving historically, one would expect a mixture of hierarchical relations and heterarchical positions.

Herbert Simon (1969; 1973) showed that such a complex system can be expected to remain *nearly* decomposable. Historical relations constrain the evolutionary dynamics, while the latter tend to control further development. The system develops in the present, but with reference to its history. Reflexively, it is able to restructure itself at the supra-individual level, that is, as an unintended effect (Leydesdorff, 2002 and 2005). Nearly decomposable systems are not crisp, but remain fuzzy: different sets can be partial subsets of one another (Bradford, 1934; Leydesdorff & Bensman, forthcoming). For example, the journals of American professional associations function as elite institutions across cognitive delineations among specialties (Bensman, 1996). The social and the intellectual organization of the sciences can be expected to interact (Withley, 1984; Leydesdorff, 1995).

Because of the interwovenness of different organizing principles, the decomposition remains sensitive to the choices of the various parameters involved, such as the seed journal(s) for collecting a citation environment, the threshold levels, similarity criteria, and the clustering algorithm. In other words, the vectors of the citation distributions among journals span a multi-dimensional space in which clouds can be distinguished, but delineation of these clouds remains fuzzy at the edges (Bensman, 2001) and varies with the perspectives chosen by the analyst (Leydesdorff & Cozzens, 1993; McKain, 1991). However, if there are no privileged positions, then one can also leave the choice of a perspective to the end-user.

The aggregated journal-journal citation matrix can be reconstructed from the information contained in the *Journal Citation Reports* of the *(Social) Science Citation Index*. Given the availability of the Internet and user-friendly visualization programs (like Pajek), the user is able to choose a point of access to the landscape of journals from the position of any of the journals involved. Let me demonstrate this visualization technique below by pursuing the analysis for this journal—*Environment and Planning B – Planning and Design*—as the point of access to its own citation environment. In the final section I return to the new possibilities which such an approach provides, for example, in evaluating research.

*Environment and Planning B* as a journal is an interesting case for a number of reasons. The editorial statement of this journal specifies that the journal wishes to "become a forum for major research in the application of computers to planning and design," and the promotion of "new approaches to planning and design which reflect formal methods or inquiry and analysis." The knowledge base of these efforts itself provides a landscape which reflects the relevant environments of the journal as a specific point of entry to these environments. Unlike the physical environment, the virtual environment is multi-dimensional and therefore may allow for relationships along dimensions other than the ones which have grown historically. The objective of this special issue is to explore these relations and a reflection on the position of the journal itself within these relations thus suits the purpose.

**Methods and materials**

The data were harvested from the *Journal Citation Report* of the *Science Citation Index* and the *Social Science Citation Index 2004.* The two indices processed 5,968 and 1,712 source journals, respectively, during the year under study.[2] The citation indices can be considered as huge matrices in which the cited journals provide information for the row vectors and citing journals for the column vectors (or vice versa); the cell values are equal to the number of unique citation relations at the article level. These matrices are asymmetrical, and the main diagonal—representing "within-journal" citations—provides an outlier in the otherwise skewed distributions. Within-journal citation traffic accounts for about 10% of the total citation traffic.[3]

---

[2] Not all source journals are processed actively, that is, in the citing dimension. In the *Science Citation Index,* 192 journals were not actively processed in this dimension. These journals are only registered when cited by other journals. In the *Social Science Citation Index* the corresponding number of inactive journals was forty (Leydesdorff, 2005b).

*Environment and Planning B* is included in the *Social Science Citation Index.* The journal was cited in 2004 a total of 506 times, and the aggregate of citations on the pages of the journal itself adds up to 1,763 times. Of these citations, 102 are within-journal "self"-citations. The impact factor is 0.495 because this indicator only takes citations to papers in the last two years into account (Garfield, 1979; Monastersky, 2005), while my figures refer to the "total cites." The impact factor can be considered as an indicator of the research front, while "total cites" is perhaps a better indicator of a journal's prestige (Bensman, forthcoming; Leydesdorff, forthcoming).

---

[3] The ISI aggregates all single citation relations under a category "all others." In this analysis, these missing values are not further considered.

|                     |     |     |     |     |     |     |     |     |     | →   | *citing* |
|---------------------|-----|-----|-----|-----|-----|-----|-----|-----|-----|-----|----------|
| *Cities*            | 38  | 16  | 7   | 0   | 10  | 0   | 0   | 0   | 2   | 28  | 101      |
| *Environ Plann A*   | 11  | 228 | 15  | 8   | 32  | 0   | 0   | 13  | 5   | 111 | 423      |
| *Environ Plann B*   | 2   | 25  | 102 | 34  | 24  | 0   | 0   | 3   | 0   | 22  | 212      |
| *Int J Geogr Inf Sci* | 0 | 14  | 13  | 82  | 0   | 0   | 0   | 2   | 0   | 0   | 111      |
| *J Am Plann Assoc*  | 2   | 6   | 8   | 0   | 60  | 3   | 0   | 4   | 0   | 18  | 101      |
| *J Archit Plan Res* | 0   | 0   | 7   | 0   | 6   | 9   | 0   | 0   | 0   | 0   | 22       |
| *J Urban Plan D-Asce* | 2 | 7   | 10  | 3   | 15  | 0   | 3   | 0   | 0   | 5   | 45       |
| *Prof Geogr*        | 3   | 16  | 6   | 13  | 10  | 0   | 0   | 69  | 2   | 14  | 133      |
| *Prog Plann*        | 13  | 34  | 7   | 21  | 13  | 0   | 0   | 3   | 7   | 33  | 131      |
| *Urban Stud*        | 23  | 73  | 14  | 2   | 35  | 2   | 0   | 7   | 5   | 250 | 411      |
| ↓ *cited*           | 94  | 419 | 189 | 163 | 205 | 14  | 3   | 101 | 21  | 481 | 1690     |

**Table 1**: citation matrix of ten journals citing *Environment and Plannning B* to more than 1% of its total citations within the *Social Science Citation Index 2004*.

Table 1 shows the citation matrix for the ten journals that cited *Environment and Planning B* more than five times in 2004, that is, above one percent of its total citation rate of 506. I shall use the threshold of one percent throughout this study, but in principle this parameter can be changed (He & Pao, 1986).

As the similarity measure between the distributions for the various journals included in a citation environment, I use the cosine between the two vectors or, in other words, the geometrical mean. Unlike the Pearson correlation coefficient, the cosine does not normalize for the arithmetic mean (Jones & Furnas, 1987; Ahlgren *et al*., 2003). One advantage of this measure is its further development into the so-called vector-space model for the visualization (Salton & McGill, 1983).

The resulting cosine matrices for all journals included in the *Social Science Citation Index* were brought online at http://users.fmg.uva.nl/lleydesdorff/jcr04s/cited and http://users.fmg.uva.nl/lleydesdorff/jcr04s/citing, respectively, in a format which allows the user to generate a map for each journal using Pajek. Pajek is a visualization program made freely available for non-commercial use at http://vlado.fmf.uni-lj.si/pub/networks/pajek .

At the indicated pages one is able to scroll down the list of journals, for example, to *Environment and Planning B*. Clicking on the journal's name produces the following file:

```
*Vertices 10
1 "Cities" 0.0 0.0 0.0 x_fact 3.313609 y_fact 5.562130
2 "EnvironPlannA" 0.0 0.0 0.0 x_fact 11.301775 y_fact 24.792899
3 "EnvironPlannB" 0.0 0.0 0.0 x_fact 5.147929 y_fact 11.183432
4 "IntJGeogrInfSci" 0.0 0.0 0.0 x_fact 4.792899 y_fact 9.644970
5 "JAmPlannAssoc" 0.0 0.0 0.0 x_fact 8.579882 y_fact 12.130178
6 "JArchitPlanRes" 0.0 0.0 0.0 x_fact 0.295858 y_fact 0.828402
7 "JUrbanPlanDAsce" 0.0 0.0 0.0 x_fact 0.000000 y_fact 0.177515
8 "ProfGeogr" 0.0 0.0 0.0 x_fact 1.893491 y_fact 5.976331
9 "ProgPlann" 0.0 0.0 0.0 x_fact 0.828402 y_fact 1.242604
10 "UrbanStud" 0.0 0.0 0.0 x_fact 13.668639 y_fact 28.461538
*Matrix
0.000000 0.458386 0.216525 0.000000 0.481792 0.000000 0.000000 0.000000 0.693832 0.645169
0.458386 0.000000 0.296918 0.216815 0.568138 0.000000 0.000000 0.277381 0.713839 0.674903
0.216525 0.296918 0.000000 0.500868 0.481952 0.000000 0.000000 0.000000 0.000000 0.273297
0.000000 0.216815 0.500868 0.000000 0.204519 0.000000 0.000000 0.205004 0.232916 0.000000
0.481792 0.568138 0.481952 0.204519 0.000000 0.373303 0.000000 0.286081 0.536347 0.633483
0.000000 0.000000 0.000000 0.000000 0.373303 0.000000 0.000000 0.000000 0.000000 0.204961
0.000000 0.000000 0.000000 0.000000 0.000000 0.000000 0.000000 0.000000 0.000000 0.000000
0.000000 0.277381 0.000000 0.205004 0.286081 0.000000 0.000000 0.000000 0.353497 0.222618
0.693832 0.713839 0.000000 0.232916 0.536347 0.000000 0.000000 0.353497 0.000000 0.735139
0.645169 0.674903 0.273297 0.000000 0.633483 0.204961 0.000000 0.222618 0.735139 0.000000
```

**Table 2**: Input file of the citation environment of *Environment and Planning B* in 2004 in the format used by Pajek for the visualization.

This input file for Pajek (Table 2) first provides the names of the journals with a number of parameters. The names of the journals are made available as labels for the nodes. The meaning of the other parameters will be discussed in a later section. The matrix is a symmetrical matrix of the cosine values. Values of the cosine below 0.2 are suppressed in order to enhance the visible patterns in the map.

**Results**

The file provided in Table 2 can be saved as a text file and then be read directly into Pajek. This results in the picture shown in Figure 1.

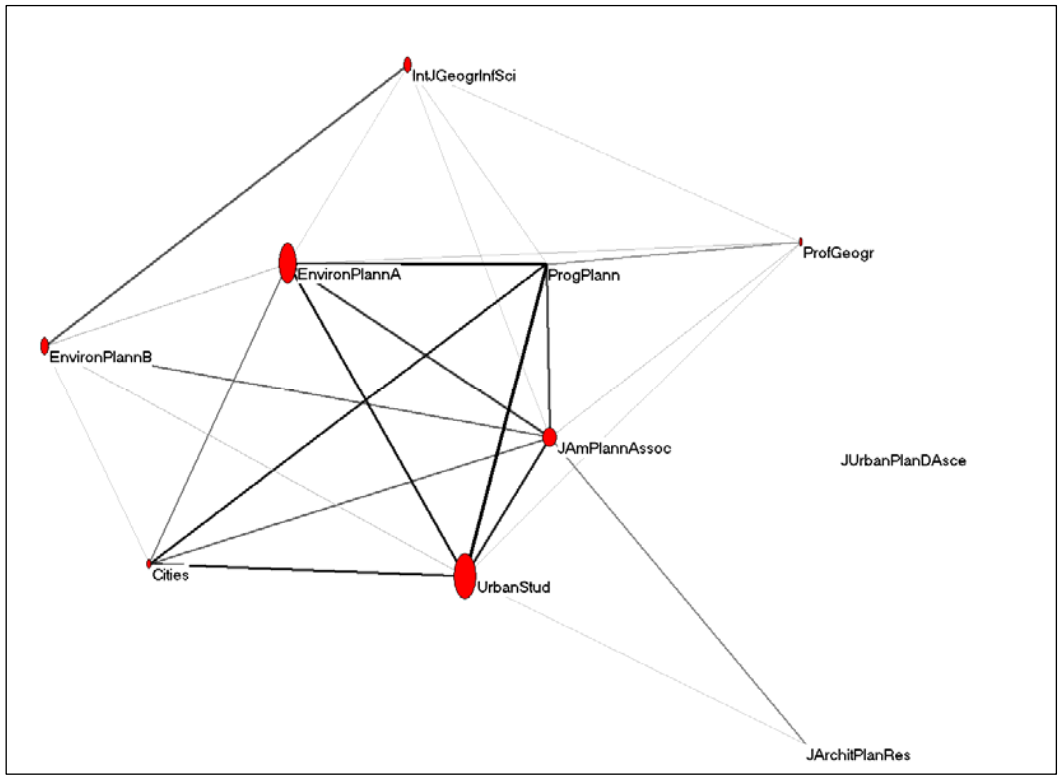

**Figure 1**: Citation impact environment of *Environment and Planning B* in 2004.

The colour and size of the nodes and the grey-shading of the links can be chosen as options within Pajek. Additionally, various cluster algorithms based on graph theory are available for the partitioning. Let us first focus on the graphs and return to the relative sizes of the nodes thereafter.

The picture of Figure 1 shows a densely connected core set of five journals focussing on planning. The *Journal of the American Planning Association* occupies a somewhat different position because it is more closely related with *Environment and Planning B* than the other journals. This journal shows less similarity in its citation patterns with the other journals in the central set, but its citation pattern is also associated with the *International Journal of Geographical Information Science.* We shall see below that these two journals function more than the others at the interface with the natural science environment, while the focus is here on the domain of the *Social Science Citation Index.*

**Rotated Component Matrix(a)**

|  | Component | | | | |
| --- | --- | --- | --- | --- | --- |
|  | 1 | 2 | 3 | 4 | 5 |
| PROG PLANN | .863 |  | -.149 | .129 | .114 |
| CITIES | .732 | -.154 | -.304 | -.359 |  |
| URBAN STUD | .722 | .419 |  |  | .144 |
| ENVIRON PLANN A | .664 | .367 | .171 | .194 |  |
| J AM PLANN ASSOC | .123 | .888 |  |  |  |
| ENVIRON PLANN B | -.145 | .188 | .812 | -.172 |  |
| INT J GEOGR INF SCI | -.132 | -.572 | .679 |  | .140 |
| PROF GEOGR |  |  | -.114 | .940 |  |
| J URBAN PLAN D-ASCE | -.251 |  | -.143 | -.110 | -.917 |
| J ARCHIT PLAN RES | -.496 | .251 | -.497 | -.264 | .523 |

Extraction Method: Principal Component Analysis. Rotation Method: Varimax with Kaiser Normalization.
a Rotation converged in 10 iterations.

**Table 3**: Factor analysis of cited patterns in the citation environment of *Environment and Planning B*.

These visual results are validated by a factor analysis of the above citation matrix using SPSS. The factor analysis shows the interfactorial position of the *Journal of the American Planning Association* as Factor 2, and the different positions of the two journals which load on Factor 3. The three remaining journals can be considered as isolates which span other (but smaller) dimensions of this citation matrix.

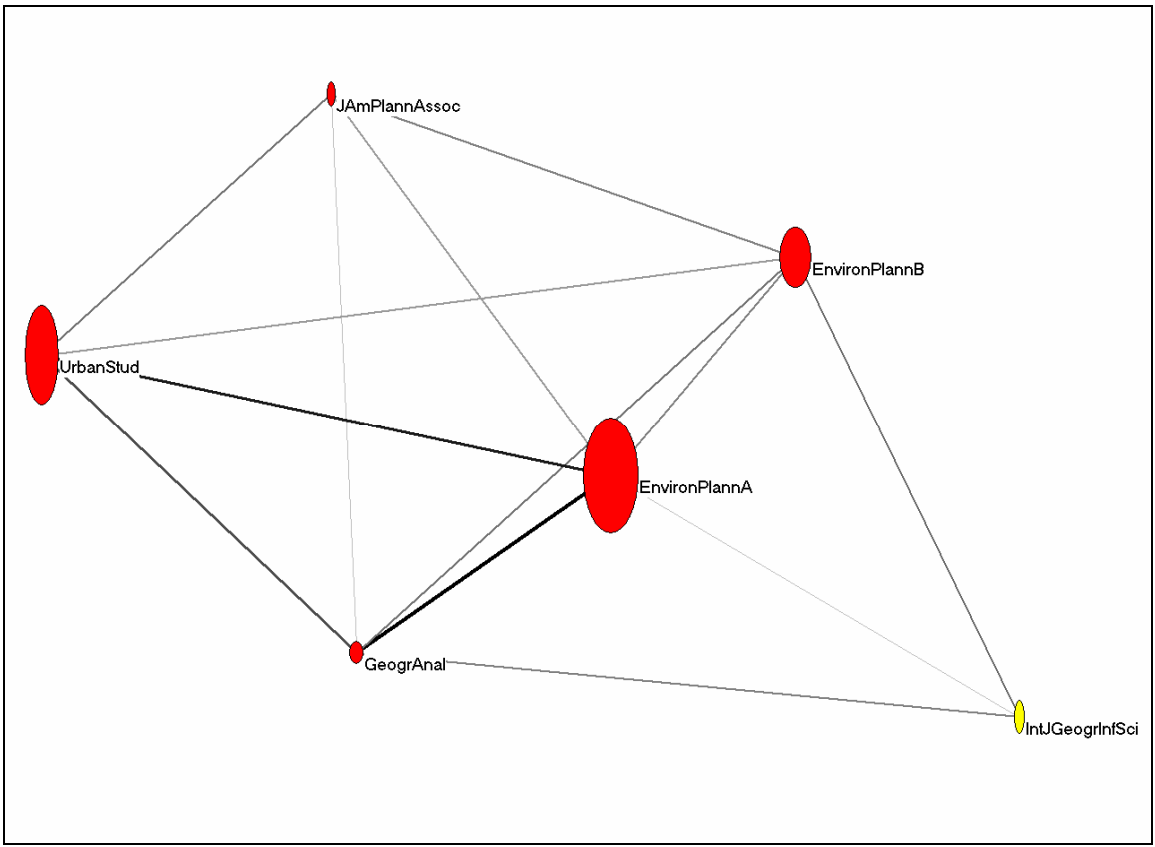

**Figure 2**: Citing patterns of journals in the citation environment of *Environment and Planning B*.

In Figure 2, the corresponding threshold in the citing environment of *Environment and Planning B* is used. Since the aggregate of the articles in this journal provided 1,763 references, one percent corresponds with 18 or more citations. Only six journals meet this threshold value, to a total of 225 citations. (Remember that 102 of these citations are within-journal self-citations of *Environment and Planning B*.) The tail of the distribution in the citing dimension is extremely large (as usual): 937 journals are cited only once. In other words, authors in the journal reach out to a large number of sources (Liu, 2005).

Nevertheless, the graph is very clear and well connected. *Environment and Planning B* is part of the core graph more than in the cited dimension, but it has retained its preferential relation with the *International Journal of Geographical Information Science*. *Geographical Analysis* has joined the core set, while it was not central in the cited dimension.

Let us now turn to the shape of the nodes and their differences in size. In the input text file (Table 2 above), the ten journals in the citation environment of *Environment and Planning B* were first defined as vertices with a label. Thereafter, three parameters are available in Pajek for fixing the coordinates of the nodes in the x, y, and z-direction (which I did not use in this study). The two parameters "x_fact" and "y_fact" provide a value for the magnification of the node in the two main directions. (Other parameters can be added, for example, in order to change the shape of the nodes from circles and ellipses into boxes or diamonds, and to control for the interior color of the nodes.)[4]

In this design, I use the two parameters of the size of the node to indicate the percentage contribution to the thus selected citation environment both including and excluding within-journal citations. For example, *Environment and Planning B* was cited within this environment 189 times, of which 102 were within-journal citations. The total number of citations in the citation matrix among these ten journals—the grandsum ($N = \sum c_{ij}$)—is 1,690 and thus, the percentage of the citations obtained by *Environment and Planning B* within this environment is (189/1,690) * 100 = 11.18%. This percentage is conveniently used as the value of the parameter "y_fact." After correction for within-journal citations, the percentage become ((189-102)/1.690) * 100 = 5.15%. This value is used analogously for the parameter "x_fact." The ASCII file provides both the inputs needed for drawing the picture in Pajek and numerical information about these percentages for users who are interested quantitatively in the local impact factors of journals in specific citation environments.

---

[4] One should be aware that the information contained in these visualization parameters will be lost if the Pajek-files are subsequently exported for the purpose of further processing and statistical analysis in programs like NetDraw or UCINET.

The aspect ratio of each node in the map reflects the self-citing nature of each journal: taller nodes represent journals with large self-citation fractions, while rounder nodes indicate lower self-citation fractions. The local impact factors are expressed as percentage shares of the grandsum of the citation environment, since the use of percentages makes the sizes independent of the citation characteristics of the specialties under study. Note that the within-journal citation rate in any year is a constant for each journal. However, the weight of this constant in each environment ($N = \sum c_{ij}$) and in the total number of citations of the journal ($\sum c_i$) varies with the environment and thus with the choice of the seed journal. In other words, the shapes and sizes of the nodes are environment-dependent.

**Extension to the *Science Citation Index***

The ISI dataset is organized in two separate indices, that is, for the sciences and the social sciences, respectively. Most journals can be evaluated using one of these two domains, but in the case of *Environment and Planning B* this is debatable. Although its citing patterns are focused on journals in the social siences, the articles in this journal are also cited by journals in the information and computer sciences. The latter are often included in the *Science Citation Index.*

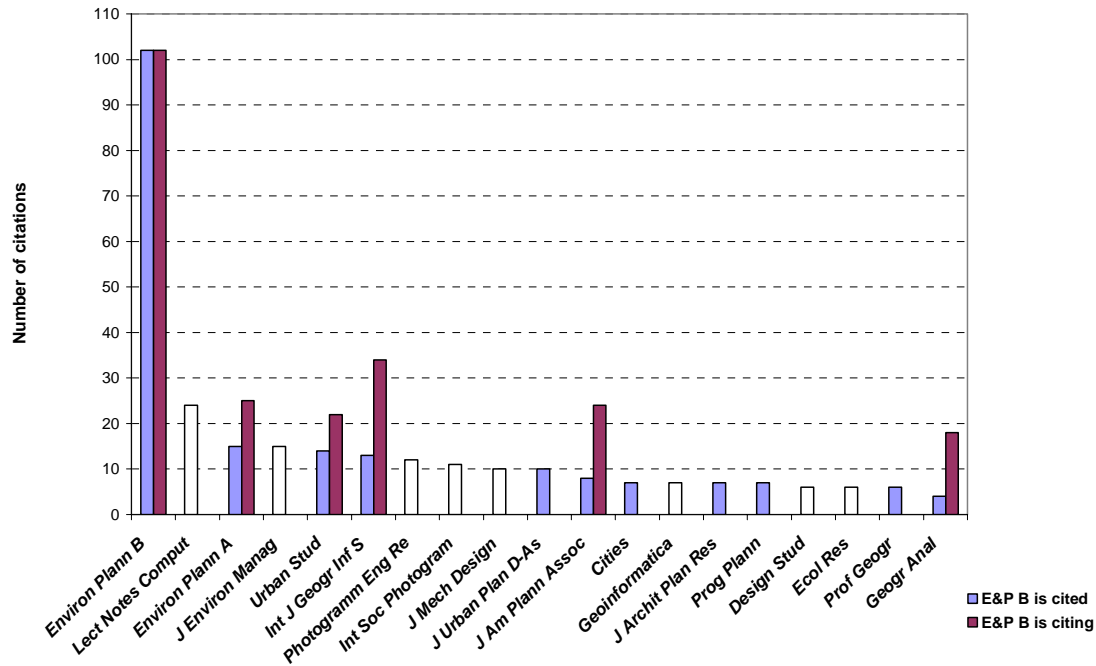

**Figure 3**: Citation distribution of journals citing and being cited by *Environment and Planning B* in 2004. The journals indicated in white are not included in the *Social Science Citation Index*, but in the *Science Citation Index*.

Figure 3 shows that in addition to the ten journals discussed in the cited dimension above, eight journals included in the *Science Citation Index* cite articles from *Environment and Planning B* to an extent of more than one percent of its total rate (506). For the purpose of this study, I merged the two databases. The journals added from the *Science Citation Index* will be indicated below with diamonds instead of ellipses. Figure 4 provides the picture in the cited dimension. As noted, the citing dimension remains unchanged because articles in *Environment and Planning B* cite mainly from journals in the social science domain.

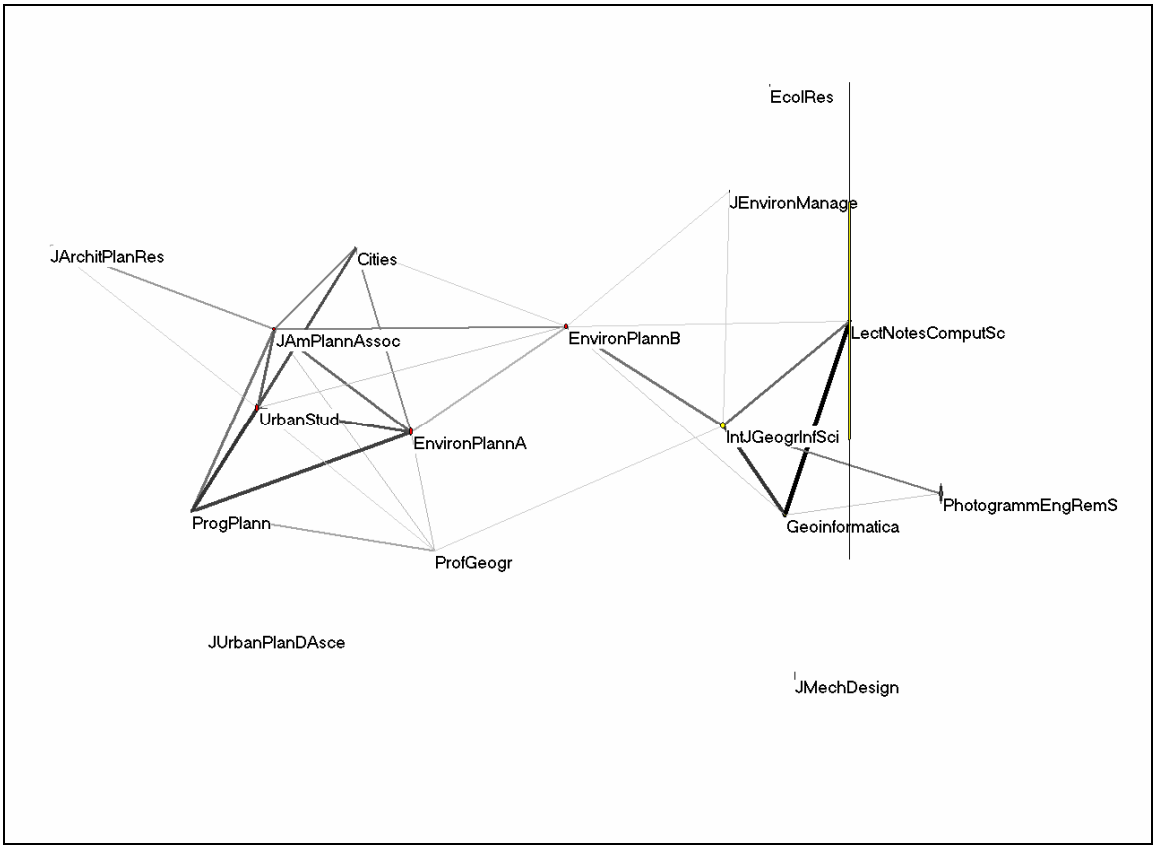

**Figure 4**: Citation impact environment of *the Environment and Planning B* in the combined *Science Citation Index* and *Social Science Citation Index* 2004.

This figure shows the different role *Environment and Planning B* and the *International Journal of Geographic Information Science* played in the above analysis (Figure 1) and the corresponding factor analysis (Table 2). *Environment and Planning B* functions almost as a so-called articulation point in this graph: it is part of two graphs, one in the social and another in the natural sciences. Through its linkage with the *Lecture Notes in Computer Science (LNCS),* however, a journal of a different order of magnitude is included in this citation environment. The size of this journal and the volume of its citations makes it impossible to see more details about the differences in citation impacts among the smaller journals.

*LNCS* has a major citation rate (32,749 citations), but not in this specific environment. The large vertical line indicates the huge value on the diagonal of "within-journal" citation rates (18,005). However, *LNCS* is not a normal journal, but rather a collection of special-topic issues on a wide variety of subjects in the computer sciences. In these special-topic issues authors are often encouraged and inclined to cite one another. However, it is easy to remove this artifact because the input files can be edited. In Figure 5, I use the default value for the two parameters which control the size. Such a correction enables us to inspect the local citation impact of the other journals involved. Figure 5 exhibits otherwise the same configuration as Figure 4, but the vertices were enlarged proportionally in order to visualize the structure among them.

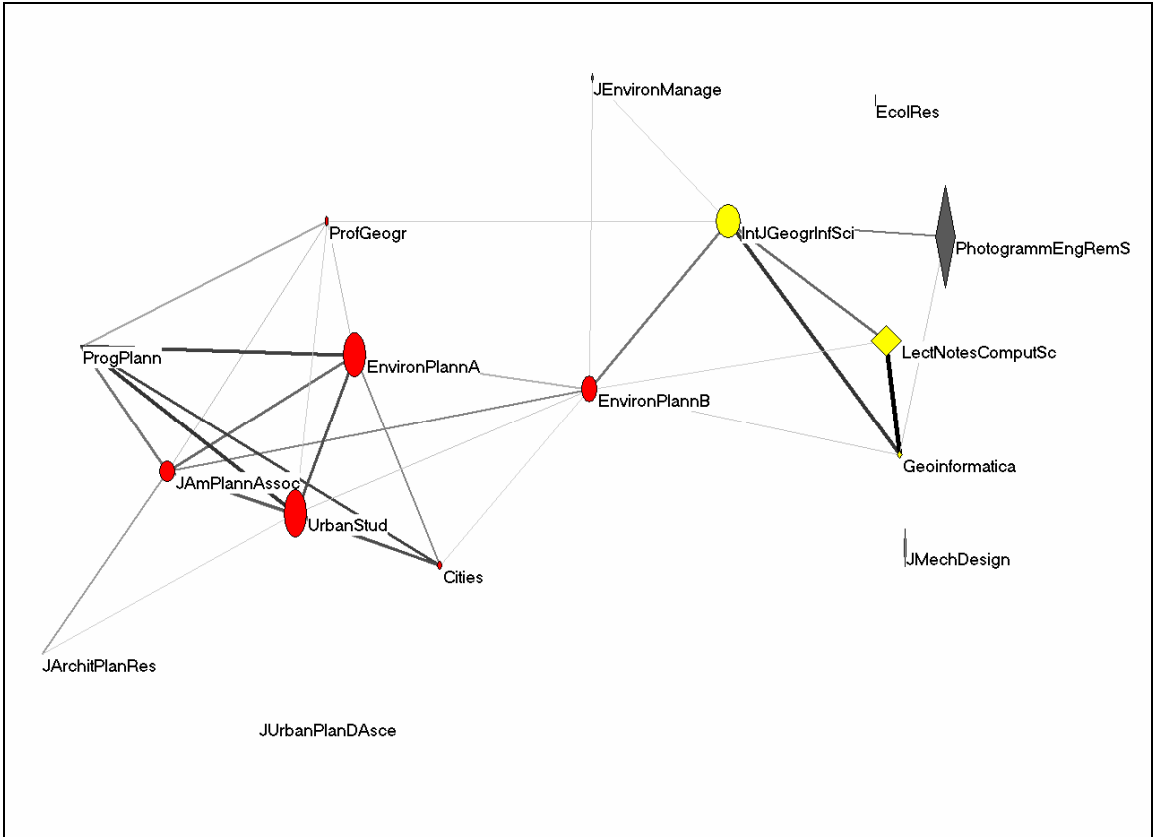

**Figure 5**: Local citation impacts of journals in the citation environment of *Environment and Planning B* in 2004. (*Social Science Citation Index* and *Science Citation Index* combined.)

Using the *k*-core clustering algorithm (available in Pajek), *Environment and Planning B* is attributed to the group of social science journals in this environment. It functions also as a window which makes results from the social sciences available to the information sciences (which are grouped on the right side of the picture). However, the *International Journal of Geographical Information Science* is attributed to the cluster on the other side of the interface. Within this interdisciplinary context, the relative citation impact of *Environment and Planning B* is larger than in the exclusively social-science context as analyzed above (Figures 1 and 2).

**The relation with GeoVis and InfoVis Journals**

Because of the focus of this special issue, the question was raised how these citation environments relate to journals with a focus on geographical information systems and information visualization. Information science and technology is another field at the interface between the social sciences and the sciences (Leydesdorff, forthcoming). The *Journal of American Society of Information Science and Technology*, for example, can be considered as a leading journals in this set. However, the relation between the two fields is visible when one uses the journal *Geoinformatics* as the seed journal. This journal entertains citation relations with 39 journals in 2004 (no threshold). Thirty-two of these journals form a graph when the threshold level of the cosine between their being-cited patterns is set at larger than and or equal to 0.2 (Figure 6).

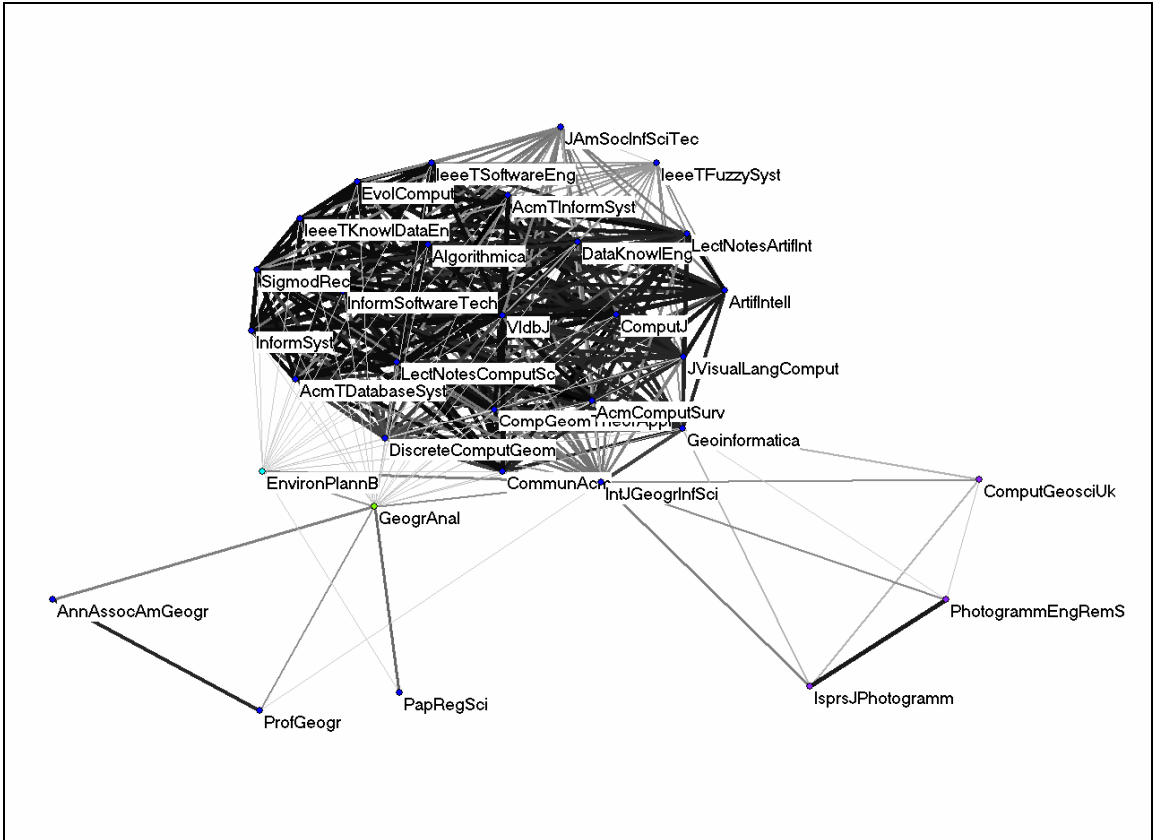

**Figure 6:** Thirty-two journals citing or cited by *Geoinformatica* more than once in 2004; being-cited patterns; cosine $\geq$ 0.2.

Central to this picture is a strong graph of computer science journals. By choosing another seed journal, one can extend this strong graph to more than hundred journals. *Environment & Planning B* and the *International Journal of Geographical Information Science* are positioned at the interface of this group with geography and geoscience journals in the lower half of the figure. The *Journal of the American Society for Information Science and Technology* can be found at the top of the graph; it is integrated in this grouping. However, from this perspective one cannot see beyond this window on the information sciences: the other information-science journals (e.g., *Scientometrics* and the *Journal of Documentation*) are not visible in this representation. In sum, both intellectual fields (geography and information science) relate through a large set of computer science journals to which they are both only marginally connected.

**Conclusions and Implications**

The evaluation of *Environment & Planning B* changes when one analyzes its local citation environment at the interface between the *Social Science Citation Index* and the *Social Science Citation Index*. The journal fullfils a specific function at the interface with the computer sciences which cannot be understood from reading the impact factors of journals which belong to its closest neighbours in terms of citation relations (Table 1).

|  | Impact factor | Total number of citations |
|---|---:|---:|
| *Cities* | 0.818 | 241 |
| *Environ Plann A* | 1.622 | 1,861 |
| *Environ Plann B* | 0.495 | 506 |
| *Int J Geogr Inf Sci* | 1.234 | 899 |
| *J Am Plann Assoc* | 0.911 | 659 |
| *J Archit Plan Res* | 0.222 | 106 |
| *J Urban Plan D-Asce* | 0.752 | 764 |
| *Prof Geogr* | 1.000 | 704 |
| *Prog Plann* | 0.200 | 72 |
| *Urban Stud* | 1.127 | 1,681 |

**Table 4**: Impact factors and total numbers of citations of ten journals citing *Environment and Plannning B* to more than 1% of its total citations within the *Social Science Citation Index 2004*.

Table 4 shows additionally that the total number of citations—which is sometimes considered as a better indicator than impact factors (see above)—is highly correlated with impact factors (Spearman's $\rho = 0.891$; $p < 0.01$).[5] Both in terms of total citations and impact factors, *Environment & Planning B* remains considerably behind its two direct neighbours on both sides of the interface: *Environment & Planning A* and the *International Journal of Geographical Information Science*.

---

[5] Spearman's $\rho$ between these two indicators is 0.746 for the *Social Science Citation Index* and 0.726 for the *Science Citation Index*. Both correlations are significant at the 1% level.

When the journal environment is extended with the *Science Citation Index*, the specific role of *Environment & Planning B* can be made visible. Unlike the *International Journal of Geographical Information Science*, *Environment & Planning B* remains part of the graph representing journals in the field of planning, but it functions as the main articulation point of this group within the computer sciences. From the perspective of the computer sciences, *Environment & Planning B* belongs to the relevant surroundings like the *Journal of the American Society of Information Science and Technology* at the other end. It shares this position with the *International Journal of Geographical Information Science*, but this latter journal can be considered as belonging to the computer sciences; it is accordingly included in both indices. Authors in *Environment & Planning B*, however, cite unambiguously from other social science journals belonging to this specialty.

In a recent article, Monasterky (2005) listed a number of problems with the ISI-impact factor and its use in the evaluation of scientific contributions. The problems involved have long been familiar to those who study science and research evaluations (Moed, 2005; Bensman, forthcoming). While impact factors are increasingly used for research evaluation, the evaluators often fail to mention that the average impact factors vary by orders of magnitude among fields and even among specialties within fields. For example, impact factors in toxicology are considerably lower than in immunology.

One solution has been suggested by Hirst (1978), who proposed introducing "discipline impact factors." More recently, Bensman (forthcoming) showed that more than with an impact factor, faculty usage and appreciation of journals correlates with the total number of citations to a journal. Citations can be considered a measure of a journal's prestige, while the impact factor follows the development of the field at the frontiers of research (Price, 1965).

It could be shown that an unambiguous clustering of the aggregated journal-journal citation matrix into disciplines and specialties is impossible (Leydesdorff, 2006). The various subsets overlap for very different reasons, such as communalities in the subject matter, methods, nationality, language, type of publisher or purpose. Each journal has its own unique environment created through the process of citing and being cited. Journals also differ in terms of their within-journal ("self"-)citation rates.

For these reasons, I have created input files that enable users—without advanced computer literacy—to generate maps of the citation neighbourhoods of all the journals included in the *Science Citation Index* and the *Social Science Citation Index* 2004. The contributions to the total number of citations in this local environment are normalized, and a local impact factor can then be computed. This local impact factor is relative to the citation density in the specific environment of interest. Furthermore, the local impact can be corrected for within-journal citations. I use the horizontal axis of the node for this corrected local impact, while the vertical axis is used for the local impact including self-citations. Thus, the nodes are represented as ellipses whose links indicate the strength of their citation relations with journals in the specifically relevant environments. Clustering algorithms are available within Pajek for colouring the visualizations differently.

The advantages of using this local impact factor are that (1) normalization over the total citations in the relevant citation environment is more indicative of the intellectual status of a given journal than an average normalized over the number of publications; (2) the evaluation can be made for each journal in the ISI set and related to the journal's specific citation environment; (3) the correction for within-journal citations is available both numerically and from the visualizations. The input text files provide the numerical information for these values as percentages. Moreover, this information is readily available on the Internet for anyone who wishes to use it.